\documentstyle[12pt,epsfig]{article}

\begin{document}

%%%%%%%%%%%%%%%%%%%%%%%%%%%%%%%%%%%%%%%%%%%%%%%%%%%%%%%%%%%%
%% Please, for the formatting just include here the standard
%% elements: title, author, date, plus TDAScode
%%%%%%%%%%%%%%%%%%%%%%%%%%%%%%%%%%%%%%%%%%%%%%%%%%%%%%%%%%%%
\title{Exploring the Gamma Ray Horizon with the next generation of
Gamma Ray Telescopes. Part 3: Optimizing the observation schedule
of $\gamma$-ray sources for the extraction of cosmological
parameters}
\author{O.Blanch, M.Martinez\\
{\it IFAE, Barcelona (Spain) }}
%\date{April 10, 2001\\ (outdated)}
%%%%%%%%%%%%%%%%%%%%%%%%%%%%%%%%%%%%%%%%%%%%%%%%%%%%%%%%%%%%

%% title %%%%%%%%%%%%%%%%%%%%%%%%%%%%%%%%%%%%%%%%%%%%%%%%%%%%
\maketitle

%% abstract %%%%%%%%%%%%%%%%%%%%%%%%%%%%%%%%%%%%%%%%%%%%%%%%%
\begin{abstract}
The optimization of the observation schedule of $\gamma$-ray
emitters by the new generation of Cherenkov Telescopes to extract
cosmological parameters from the measurement of the Gamma Ray
Horizon at different redshifts is discussed. It is shown that
improvements over 30\% in the expected cosmological parameter
uncertainties can be achieved if instead of equal-observation
time, dedicated observation schedules are applied.
\end{abstract}

%% contents %%%%%%%%%%%%%%%%%%%%%%%%%%%%%%%%%%%%%%%%%%%%%%%%%
%\thetableofcontents

\newpage

%% body %%%%%%%%%%%%%%%%%%%%%%%%%%%%%%%%%%%%%%%%%%%%%%%%%%%%%

%------------------------------------------------------------
\section{Introduction}

Imaging Air \v{C}erenkov Telescopes (IACT) have proven to be the
most successful tool developed so far to explore the $\gamma$-ray
sky at energies above few hundred GeV. A pioneering generation of
installations has been able to detect a handful of sources and to
start a whole program of very exciting physics studies. Nowadays a
second generation of more sophisticated Telescopes is starting to
provide new  observations. One of the main characteristics of some
of the new Telescopes is the potential ability to reduce the gamma
ray energy threshold below $\sim 30$ GeV \cite{MAGIC}.

In the framework of the Standard Model of particle interactions,
high energy gamma rays traversing cosmological distances are
expected to be absorbed through their interaction with the diffuse
background radiation fields, or Extragalactic Background Field
(EBL), producing $e^+ e^-$ pairs. Then the flux is attenuated as a
function of the gamma energy $E$ and the redshift $z_{q}$ of the
gamma ray source. This flux reduction can be parameterized by the
optical depth \(\tau(E,z_{q})\), which is defined as the number of
e-fold reductions of the observed flux as compared with the
initial flux at $z_{q}$. This means that the optical depth
introduces an attenuation factor \(\exp[-\tau(E,z_{q})]\)
modifying the gamma ray source energy spectrum.

The optical depth can be written with its explicit redshift and energy
dependence \cite{Stecker-96} as:

\begin{equation}
\tau(E,z) =
\int_{0}^{z}dz'\frac{dl}{dz'}\int_{0}^{2}dx \,
\frac{x}{2}\int_{\frac{2m^{2}c^{4}}{Ex(1+z')^{2}}}^{\infty}
d\epsilon\cdot n(\epsilon,z') \cdot \sigma[2xE\epsilon(1+z')^{2}]
\label{eq:OpD}
\end{equation}

where $x \equiv 1-\cos\theta $ being $\theta$ the angle between the
photon directions, $\epsilon$ is the energy of the EBL
photon and $n( \epsilon ,z')$ is the spectral density at the given z'.

For any given gamma ray energy, the Gamma Ray Horizon (GRH) is
defined as the source redshift $z$ for which the optical depth is
$\tau(E,z) = 1$.

In a previous work \cite{part1}, we discussed different
theoretical aspects of the calculation of the Gamma Ray Horizon,
such as the effects of different EBL models and the sensitivity of
the GRH to the assumed cosmological parameters.

Later, on \cite{part2} we estimated with a realistic simulation
the accuracy in the determination of the GRH that can be expected
from an equal-time observation of a selection of extragalactic
sources. The results obtained in that previous study
assumed an observation schedule of equal observing time per source
which was set to a canonical value of 50 hours (rather standard
assumption in IACTs). Although the actual observing time per
source might have a lot of constraints (such as significance of
the observation, physics interest of the source, competition in
time for the observation of other sources, etc...), in this work
we want to explore, taking into account just the unavoidable
observability constraints, which time scheduling would optimize
the power of this method to extract cosmological parameters and by
how much the measurement of these parameters could improve.

One must also take into account that the determination of the
cosmological parameters extensively discussed in \cite{part2} is
based on the observations of Active Galactic Nuclei (AGN), which
are intrinsically variable. Therefore one of the main parameters
to decide which AGN is observed at any time will be their flaring
state. For some AGN, it is possible to estimate its activity from
observations in other wavelength, for instance using the X-ray
data \cite{ASMWeb} provided by the All-Sky Monitor (ASM)
\cite{Levine-96} onboard the Rossi X-ray Timing Explorer.
Unfortunately, there are a lot of AGNs for which there is not
online data that would allow to infer the flaring state. Actually
in the current catalogue of sources that are monitored by ASM only
3 of the 22 used on these studies appear. So that, here we'll
present an observational scheduling for the AGNs in table
\ref{tab:GRHvsTime}, which does not care about the activity of the
source, and just optimizes the observation time to get the best
precision on the cosmological parameter measurements.\par

The work is organized as follows: in section 2, the expected
improvement in the precision of the GRH determination as a
function of the observation time is discussed. Section 3 deals
with the observational constraints in the optimization procedure.
In section 4 we describe the optimization technique employed and
describe the actual algorithm used. Section 5 presents the results
of the optimization procedure in different scenarios considered
and finally in section 6 we summarize the conclusions of this
study.

\section{Gamma Ray Horizon energy precision}

To optimize the observation time, the first step
is the study of the GRH precision as a function of the time that
is dedicated for each source. The estimated precision on the GRH
comes from the extrapolation of the detected spectra of each
source by MAGIC (see \cite{part2}). In there, the observation time
enters as a multiplicative term to get the number of
$\gamma$s.\par

\begin{figure}[ht]
  \begin{center}
    \includegraphics[height=22pc]{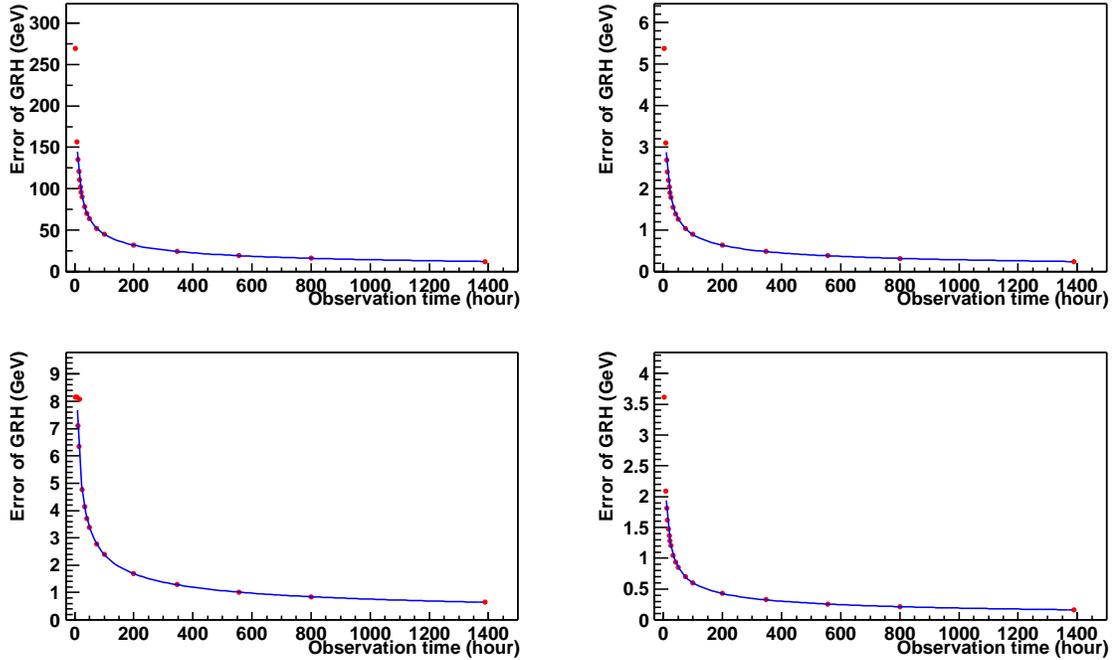}
  \end{center}
  \vspace{-0.5pc}
  \caption[Statistic error of GRH vs observation Time]
    {Evolution of the statistic precision of the GRH determination
    as a function of observation time for four of the used AGNs
    (3EG J1426+428, 3EG J1255-0549, 3EG J0340-0201 and 3EG
    J1635+3813). The blue line is the fit to one over square-root of time.}
\label{fig:GRHvsTimeStat}
\end{figure}

In figure \ref{fig:GRHvsTimeStat}, the expected $\sigma$ of the
GRH ($\sigma_{grh}$) using several observation times is shown. In
these plots only the statistical error from the fitting parameter
is shown. That error comes from the error bars in the extrapolated
spectra. On the one hand there is the uncertainty on the flux
($\Phi$) that is modeled as ``n'' times the square-root of $\Phi$,
which is proportional to $\sqrt{N_{\gamma}}$ being $N_{\gamma}$
the number of detected $\gamma$s. On the other hand, the error on
the determination of the energy improves also with
$\sqrt{N_{\gamma}}$ if one assumes a gaussian statistic.
Therefore, one expects that the $\sigma_{grh}$ decreases with the
square-root of time. Actually, the extrapolated errors show a good
agreement with the blue line that is a fit to:

\begin{equation}
\sigma_{grh}= k / \sqrt{time} \label{eq:GRHvsTimeStat}
\end{equation}

\begin{figure}[ht]
  \begin{center}
    \includegraphics[height=22pc]{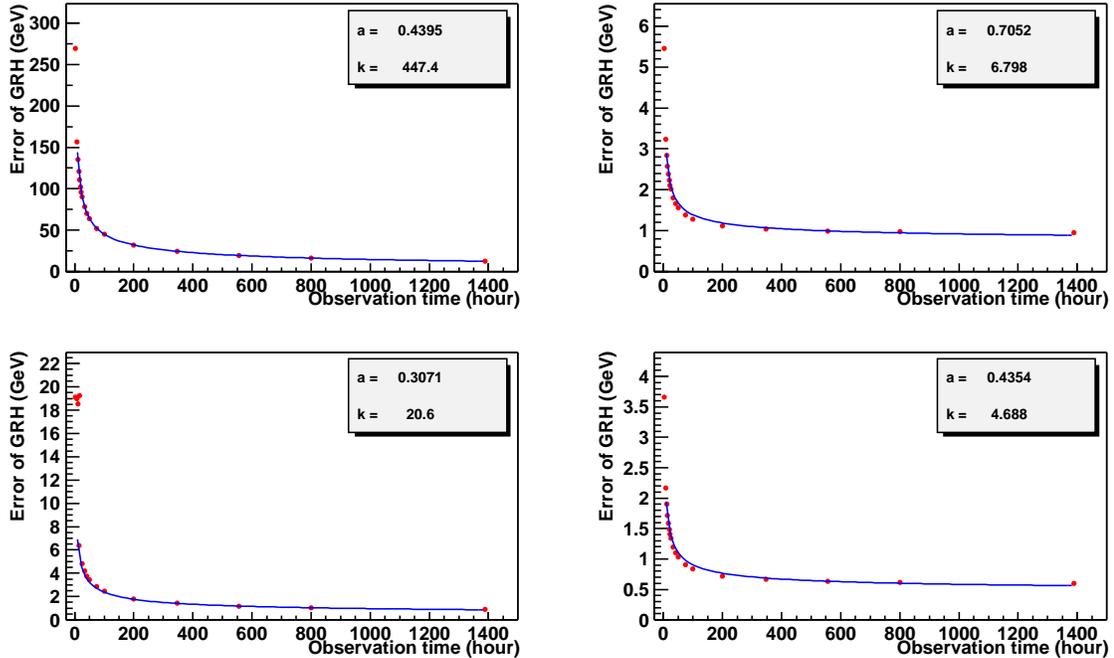}
  \end{center}
  \vspace{-0.5pc}
  \caption[Systematic error of GRH vs observation Time]
    {Evolution of the precision of the GRH determination , adding
    the systematics due to the approximation in the fit of the spectra,
    as a function of observation time for
    four of the used AGNs (3EG J1426+428, 3EG J1255-0549, 3EG J0340-0201 and
    3EG J1635+3813). The blue line is the fit to equation
    \ref{eq:GRHvsTimeSys}.}
\label{fig:GRHvsTimeSys}
\end{figure}

This latter expression would mean that the $\sigma_{grh}$ can be
as small as desired if enough observation time is used. But it
does not represent the reality. One should also take into account
the systematics, which become more and more important when
reducing the statistical error. As it has already been discussed
in \cite{part2} the main systematic errors in the GRH
determination from the simulated experimental data are due to the
uncertainty in the global energy scale and to some approximations
used to fit the data. The former is a global systematic, which is
absolutely independent and uncorrelated to the observation time,
and hence it is not considered here. Instead, the latter should
have an impact in the precision of the GRH determination as a
function of the observation time that may be different for each
source. The main effect of those approximations is that the value
of the GRH differs slightly from the one that has been introduced.
This difference is added quadratically to the statistical error to
account for the systematic difference. Then the figure
\ref{fig:GRHvsTimeStat} has been repeated and the result for the
same 4 sources are shown in figure \ref{fig:GRHvsTimeSys}. In that
scenario the curve is fitted to :

\begin{equation}
\sigma_{grh}= a + k / \sqrt{time} \label{eq:GRHvsTimeSys}
\end{equation}

where a is the contribution coming from the systematic, which does
not decrease with the amount of observation time and therefore
becomes important when the time is large. In fact the
parameterization that have been finally used is :

\begin{equation}
\sigma_{grh}= a + GRH(50 h) * \sqrt{\frac{50 \,
hours}{time(hour)}} \label{eq:GRHvsTime}
\end{equation}

where $GRH(50 h)$ are the statistic errors for the GRH using 50
hours of observation time and $a$ is the
constant term of the above mentioned fit (table
\ref{tab:GRHvsTime}).\par

\section{Constraints}

The aim of this work is the optimization of the time dedicated to
each source, to understand which improvement can be obtained in
the cosmological measurements. Nevertheless, that time should make
sense in the frame of the possible observations performed by an
IACT such as, for instance, MAGIC. Therefore some constraints
should be set.\par

The first constraint is the total amount of time used for those
observations. For that, we used 1000 hours to compare it with the
``50 h per source'' configuration. On that naive configuration 50
hours were chosen since it is a reasonable time to spend in a
single source and it was already a criteria to do the list of the
best MAGIC targets in \cite{flix-ICRC03}. Since 20 sources were
used (see reference \cite{part2}), it accounts for \mbox{1 000
hours}. Moreover, taking into account that \v{C}erenkov Telescopes
have typically observations times of about \mbox{1 200 hours} per
year, the limit used could be reached even in one single year. And
it is more than acceptable for 2-3 years, since AGNs are one of
the main targets of the new generation of IACTs.\par

One of the singularities of the astrophysics field respect to
other physics disciplines is that it studies phenomena that cannot
be generated by the humans in a laboratory. Therefore one has to
use what nature provides. In this sense, IACTs cannot observe one
given source for an infinite time during a year, not even those
1200 hours of observation time, since each source is only visible
during some months every year. Based on that fact, we have
computed the amount of time that each source is visible below 45
degrees zenith angle form the MAGIC location. In table
\ref{tab:GRHvsTime}, one can see that time for each of the used AGNs, actually it holds for the
year 2005 and it may change a few percent from year to year due to
the full moon periods. To compute the optimal distribution of
observation times, the constraint ``MaxTime'' used for each source
is :

\begin{equation}
MaxTime < T ( 1 \, year ) * Years * F \label{eq:timeAvailable}
\end{equation}

where $T(1 \, year)$ are the number of hours stated in table
\ref{tab:GRHvsTime}. ``$Years$'' is the number of years during
which data would be collected and it is set to 3. And F is the
fraction of the available time during which data would be taken.
It is set to 0.25 and it accounts for bad weather conditions, off
data needed for the standard ``On-Off'' analysis and time
dedicated to other sources or targets of opportunity.

\begin{table}[t]
\begin{center}
\fontsize{9}{10} \selectfont
\begin{tabular}{|c|cccc|}
\hline
Source Name & z & $a (GeV)$ & $GRH(50h) (GeV)$ & $T(1year) (hour)$ \\
\hline \hline
W Comae , 3EG J1222+2841 & 0.102 & 0.82 & 23.9 & 370 \\
         3EG J1009+4855 & 0.200   & 28.3 & 2.6 & 428 \\
OJ+287 , 3EG J0853+1941 & 0.306 & 3.15 & 9.1 & 453 \\
4C+15.54 , 3EG J1605+1553 & 0.357 & 21.4 & 7.7 & 403 \\
        3EG J0958+6533 & 0.368 & 0.52 & 15.2 & 317 \\
        3EG J0204+1458 & 0.405 & 0.12 & 12.7 & 404 \\
        3EG J1224+2118 & 0.435 & 0.07 & 21.6 & 408 \\
3C 279  , 3EG J1255-0549 & 0.538 & 0.69 & 1.3 & 244 \\
        3EG J0852-1216 & 0.566 & 0.73 & 2.6 & 125 \\
4C+29.45 , 3EG J1200+2847 & 0.729 & 5.02 & 3.2 & 427 \\
CTA026  , 3EG J0340-0201 & 0.852 & 0.29 & 1.76 & 316 \\
3C454.3 , 3EG J2254+1601 & 0.859 & 1.88 & 0.73 & 432 \\
        3EG J0952+5501 & 0.901 & 1.16 & 7.8 & 410 \\
        3EG J1733-1313 & 0.902 & 1.54 & 4.8 & 122 \\
OD+160  , 3EG J0237+1635 & 0.940  & 16.5 & 1.0 & 405 \\
        3EG J2359+2041 & 1.070 & 2.00 & 7.4 & 445 \\
        3EG J0450+1105 & 1.207 & 0.15 & 3.8 & 371 \\
        3EG J1323+2200 & 1.400 & 0.43 & 2.4 & 374 \\
        3EG J1635+3813 & 1.814 & 0.40 & 1.8 & 468 \\
1ES J1426+428  & 0.129 & 0.44 & 61.1 & 468 \\
\hline
\end{tabular}
\caption[Source characteristics for the time optimization.]
{\label{tab:GRHvsTime} Parameters used for the optimization of the
time observation dedicated to each source. The parameter $a$ is
the time independent term contributing to the $\sigma_{grh}$
(equation \ref{eq:GRHvsTime}). And ``$T(1year)$'' is the time that
the source is below 45 degrees zenith angle during 2005 at the MAGIC
location.}
\end{center}
\end{table}

\section{Time Optimization}

In reference \cite{part2} the capability of the new IACTs to
measure cosmological constants has been discussed as well as the
systematics on these measurements. There, the main emphasis was
put in the $68\%$ contour in the $\Omega_{m}-\Omega_{\lambda}$
plane, and it has been shown that it is competitive taking into
account the systematics ($15\%$ of energy scale, fit approximation
and unknown Extragalactic Background Light (EBL)) if a $15\%$
external constraint on the Ultra Violet (UV) background is used.
Under this scenario and scheduling 50 hours to each source, there
is also the possibility to fit $\Omega_{m}$ and
$\Omega_{\lambda}$. Now, we would like to optimize the
distribution of the observation time among the used sources to get
the best precision on the measurement of the cosmological
densities.\par

In order to optimize the distribution of the observation time by
requiring a minimum error in some given parameter, a technique
based upon a multidimensional constrained minimization using the
Fisher Information Matrix has been used.

The Fisher Information Matrix, is defined as

\begin{equation}
F_{ij} =  \left< \frac{-d^2 log L}{d \theta_i d \theta_j} \right>
\end{equation}

where $ L $ is the likelihood function of the measurements, $
\theta_i $ and $ \theta_j $ are fitting parameters and $ <...> $
denotes expected value. In "normal" conditions, it is the inverse
of the error matrix for the parameters $i,j$. For large samples, a
good estimator of $F$ is simply the function

\begin{equation}
F_{ij} = \frac{-d^2 log L}{d \theta_i d \theta_j}
\end{equation}

evaluated at $ \theta=\hat{\theta} $ namely, at the best fit
parameter values.

In case $ L $ could be simply approximated by a gaussian centered
at the measured values, then

\begin{equation}
F_{ij} = \sum_{k,l} (d f_k / d \theta_i) (V^{-1}_{kl}) (d f_l / d
\theta_j)
\end{equation}

evaluated at $ \theta=\hat{\theta} $. The $ i,j $ indices run over
all the fitting parameters (the cosmological parameters in our
case) and the $ k,l $ run over all the measurements (the GRH
measurements for different redshift in our case). $ V $ is the
error matrix of the measurements and $ f_k (\theta)$ is the
theoretical prediction for measurement $ k $.

External constraints on the parameters are included by adding
their corresponding Fisher Information Matrix. For instance, if
$\Omega_{\lambda}$ corresponds to parameter $ i=3 $, and we want
to include the constraint due to a measurement
$\Omega^{CMB}_{\lambda} \pm \Delta \Omega^{CMB}_{\lambda} $ we
just have to add to $ F_{ij} $ a matrix $ F'_{ij} $ with

\begin{equation}
F'_{33} = \left( \frac{\Omega^{CMB}_{\lambda}}
                  {\Delta \Omega^{CMB}_{\lambda}} \right)^2
\end{equation}

and zero in all the other matrix elements.

This way, one can compute the expected fit parabolic error for any
parameter without having actually to perform the fit. The expected
error in $\Omega_{\lambda}$ for instance would simply be
$(F^{-1})_{33}$.

Now one must minimize this quantity (or any desired function of
the fitting parameters) with respect to the observation time
expended in each source (which enters in the evaluation of $V$ and
hence, on $F$) with the relevant physical boundaries and
constraints. For that we use the mathematical approach implemented
in the code "DONLP2" developed by M.Spelucci \cite{Spellucci-98}.
The mathematical algorithm evaluates the function to be minimized
only at points that are feasible with respect to the bounds. This
allows to solve a smooth nonlinear multidimensional real function
subject to a set of inequality and equality constraints. In our
particular case:\par

\begin{itemize}
\item{Problem function}: It depends on the variable that we want
to minimize but it is always a combination of the elements of the
Fisher Information Matrix of the four dimensional fit in terms of
$H_0,\Omega_{\lambda}, \Omega_M$ and the amount of UV background
as described in \cite{part2}.

\item{Equality constraints}: The global amount of observation
time, which is set to 1000 hours.

\item{Inequality constraints}: The maximum available time for each
source.
\end{itemize}

The result of this procedure is an array providing the optimal
distribution of the observation times assuming parabolic errors,
though we have explicitly checked that for the optimal time
distribution the obtained precision on the fit parameters does not
depend sizably on the assumption of parabolic errors.
\par

\section{Results}

\begin{figure}[h]
  \begin{center}
    \includegraphics[height=20pc]{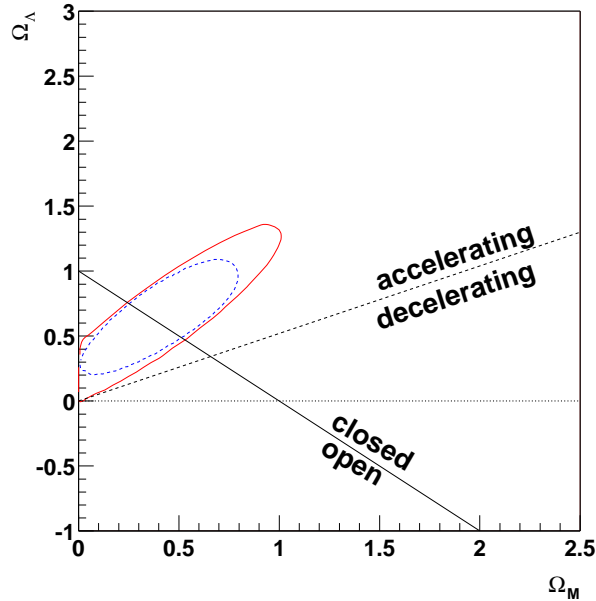}
  \end{center}
  \vspace{-0.5pc}
  \caption[Contour after time optimization]
    {Improvement on the $68\%$ contour in the $\Omega_{m} -
    \Omega_{\lambda}$ plane. The red solid line is the $68 \%$ contour,
    taking into account the systematics and imposing a $15 \%$
    constraint on the UV background, when 50 hours for each of the 20
    used sources are scheduled. The blue dashed line is the $68\%$
    contour under the same conditions but with the optimized time
    distribution.}
\label{fig:optimtime}
\end{figure}

After the optimization to get the minimum error on $\Omega_{m}$ or
$\Omega_{\lambda}$, this precision is improved by about $35 \%$
(see table \ref{tab:Sigmas}). It is worth to notice that the
obtained uncertainties for $\sigma_{\Omega_{m}}$ and
$\sigma_{\Omega_{\lambda}}$ do not significantly differ while
optimizing for one or the other. Even optimizing for the area of
$68\%$ contour in $\Omega_{m}-\Omega_{\lambda}$ plane, which is
done assuming that the contour is an ellipse, the precision
obtained is at the same level. This effect is mainly due to the
correlation between $\Omega_{m}$ and $\Omega_{\lambda}$. Therefore, we
will refer as the optimum time the one that minimizes the area of
the $\Omega_{m} - \Omega_{\lambda}$ contour. This optimum time for
each source is shown in table \ref{tab:optimtime}. One should
notice that in this table only some of the initial 20
extragalactic considered sources remain. For the others, the
optimization suggests that it is less interesting to observe them
in terms of cosmological measurements. Moreover, the remaining
sources are the ones at lowest and highest redshifts as well as
the ones with smaller errors. Both were expected to survive since
the former give the capacity to disentangle the cosmological
parameters (see reference \cite{part1}) and the latter give larger
constraints with less dedicated time. The improvement for the
$68\%$ contour can be seen in figure \ref{fig:optimtime}.\par

\begin{table}[h]
  \begin{center}
    \footnotesize
    \begin{tabular}{|l|cccc|}    \hline
      Parameter   & 50 h & $\Omega_{m}$  & $\Omega_{\lambda}$ & $68 \%
    contour$ \\ \hline
      $\sigma_{\Omega_{\lambda}}$ & 0.366 & 0.279 & 0.278 & 0.279 \\
      $\sigma_{\Omega_{m}}$       & 0.417 & 0.241 & 0.245 & 0.246 \\ \hline
    \end{tabular}
  \end{center}
  \caption[Precision of  $\Omega_{m}$  and $\Omega_{\lambda}$.]
    {Error for the cosmological densities observing for a total of
    1000 hours the considered 20 AGNs. In each column the distribution
    of these 1000 hours is done following different criteria. The
    first column is a distribution of 50 hours each source.
    Second and third are times optimized to minimize the uncertainty
    on  $\Omega_{m}$  and $\Omega_{\lambda}$. The last
    column optimizes the area of the $68 \%$ contour in the  $\Omega_{m}
    - \Omega_{\lambda}$ plane.}
  \label{tab:Sigmas}
\end{table}

\begin{table}[h]
\begin{center}
\fontsize{9}{10} \selectfont
\begin{tabular}{|c|ccc|}
\hline
Source Name & z & $ T_{area} (hour)$ & $T_{UV} (hour)$ \\
\hline \hline
W Comae , 3EG J1222+2841 & 0.102 & 60 & 278  \\
3C 279  , 3EG J1255-0549 & 0.538 & 78 & 21 \\
        3EG J0852-1216 & 0.566 & 7 & --- \\
CTA026  , 3EG J0340-0201 & 0.852 & 167 & 3 \\
3C454.3 , 3EG J2254+1601 & 0.859 & 14 & 7 \\
        3EG J0450+1105 & 1.207 & --- & 13  \\
        3EG J1323+2200 & 1.400 & 10 & 278 \\
        3EG J1635+3813 & 1.814 & 351 & 351 \\
\hline
1ES J1426+428  & 0.129 & 312 & 49 \\
\hline
\end{tabular}
\caption[Scheduled time for each source to get the best $68\%$
contour] {\label{tab:optimtime} Time scheduled for each source.
First column optimizes the area of the $68 \%$ contour and the
second one the determination of the scale factor for the UV
background.}
\end{center}
\end{table}

It has already been mentioned in \cite{part2} that a different
approach to extract information from the GRH can be done: one can
use the present constraints of the cosmological parameters to get
information on the EBL. In reference \cite{part2}, the complexity
of such analysis is discussed but a simple first step can be done
within the scenario of the 4 dimensional fit used for these
studies. If one uses the current measurements of the cosmological
parameters ( $H_{o}=72\pm4$, $\Omega_{m}=0.29\pm0.07$ and
$\Omega_{\lambda}=0.72\pm0.09$ \cite{Spergel-03}\cite{Wang-03}) as
external constraints and then optimizes the time distribution to get
the minimum error on the fourth parameter which gives a scale
factor for the UV background, one can reach a precision of $13.5
\%$. It is worth to notice that , despite the distribution of time
among them is different, the sources that are still used are
roughly the same than the ones for the  $\Omega_{m} -
\Omega_{\lambda}$ optimization.

\section{Conclusions}

In our previous works on this subject \cite{part1,part2}, it was
shown that the precision reached to measure the GRH for 50 hours
of observation time is not the same for each of the 20 considered
extragalactic sources. Moreover, it was also clear that the
sensitivity to $\Omega_{m}$ and $\Omega_{\lambda}$ was larger at
high redshift and that the capability to disentangle the
cosmological parameters is based on having measurements at low and
high redshift. Therefore, it is clear that a cleverer distribution
of the observation time would lead to better results. The
optimization of that time distribution pointed out the need of
having low redshift measurements (\mbox{3EG J1426+428} at z =
0.129) as well as others at high redshift (\mbox{3EG J1635+3813}
at z = 1.814). Together with these extreme sources, the dedication
of time at sources that reach the best precision of the GRH
(\mbox{3EG J0340-0201}) would also help to improve the
results.\par

In this work, the optimal distribution of the observation time,
taking into account scheduling constraints, has been studied by
applying a technique based upon a multidimensional constrained
minimization using the Fisher Information Matrix. The results
obtained show that a proper scheduling optimized for the
determination of the cosmological parameters could allow to reduce
by $35 \%$ the error on the determination of $\Omega_{m}$ and
$\Omega_{\lambda}$ and a notable reduction of the $68 \%$ contour
in the $\Omega_{m}-\Omega_{\lambda}$ plane.

\section*{Acknowledgments}

We are indebted to R.Miquel for his advice with the use the Fisher
Information Matrix and for providing us with M.Spelucci's DONLP2
code. We want to thank our colleagues of the MAGIC collaboration
for their comments and support.

%------------------------------------------------------------

%%% BIBLIOGRAPHY %%%%%%%%%%%%%%%%%%%%%%%%%%%%%%%%%%%%%%%%

%%>>>> Use the following if you are using BibTeX for bibliography
%\theBibliography

%%>>>> Or the following if you include here by hand your
%%>>>> bibliographic entries


\begin{thebibliography}{900}

\bibitem{MAGIC} J.A. Barrio {\it et al}, {\it The MAGIC telescope}, MPI-PhE/98-5 (1998).
\bibitem{Stecker-96} Stecker F.W., De Jager O., Space Sci.Rev.
75 (1996), 401-412
\bibitem{part1} Blanch O. and Mart\'{\i}nez M., astro-ph/0107582.
\bibitem{part2} Blanch O. and Mart\'{\i}nez M., astro-ph/0406061.
\bibitem{ASMWeb} http://heasarc.gsfc.nasa.gov/xte\_weather/
(GALEX).
\bibitem{Levine-96} A.M. Levine, Astrophys. J. {\bf 469} (1996) 33
\bibitem{flix-ICRC03} D. Bastieri {\it et al}, ICRC 03 proceedings.
\bibitem{Spellucci-98} P. Spellucci, {\sl Math. Meth. Oper. Res.} {\bf 47} (1998).
\bibitem{Spergel-03} D. N. Spergel {\it et al}, Astrophys.J.Suppl. 148 (2003) 175
\bibitem{Wang-03} X.Wang {\it et al}, Phys. Rev. D{\bf 68} (2003) 123001. 


\end{thebibliography}
\end{document}